\documentclass[11pt,letterpaper]{article}

\usepackage{amsmath}
\usepackage{amssymb}
\usepackage{bbm}
\usepackage{hyperref}
\usepackage{mathrsfs}
\usepackage{mathtools}
\usepackage{enumitem}
\usepackage{booktabs,multirow}
\usepackage[font=sf]{caption}

\usepackage{natbib}
\newcommand{\ignore}[1]{}
\newcommand{\nop}[1]{}
\newcommand*{\eg}{\textit{e.g.}}

\newcommand*{\ie}{\textit{i.e.}}

\usepackage{pgfplots}
\usepackage{pgfplotstable}
\pgfplotsset{compat=1.13}  
\usepgfplotslibrary{groupplots}
\usepackage{tikz}

\author{
  Pamela Bilo-Thomas\\
  University of Louisville\\
  \texttt{pamela.thomas.1@louisville.edu}
  \and
  Clark Hogan-Taylor\\
  Moonshot\\
  \texttt{clark@moonshotcve.com}
  \and
  Michael Yankoski\\
  University of Notre Dame\\
  \texttt{myankosk@nd.edu}
  \and
  Tim Weninger\\
  University of Notre Dame\\
  \texttt{tweninger@nd.edu}
}

\title{Pilot Study Suggests Online Media Literacy Programming Reduces Belief in False News in Indonesia}

\date{}

\begin{document}

\maketitle

\begin{abstract}
Amidst the threat of digital misinformation, we offer a pilot study regarding the efficacy of an online social media literacy campaign aimed at empowering individuals in Indonesia with skills to help them identify misinformation. We found that users who engaged with our online training materials and educational videos were more likely to identify misinformation than those in our control group (total $N$=1000). Given the promising results of our preliminary study, we plan to expand efforts in this area, and build upon lessons learned from this pilot study.
\end{abstract}


\section*{Introduction}
While the use of targeted misinformation campaigns by state actors is well documented~\citep{keller2020political,zannettou2019disinformation,bradshaw2018global}, ordinary citizens often both consume and spread (mis)information through their well-meaning online activity, which can inadvertently help to spread falsehoods to their friends and followers. Many users of online and social media systems are not aware of how misinformation is generated or spread, and thus they may unwittingly participate in the acceleration or distribution of these materials. Furthermore, the more exposed to misinformation a person becomes, the more likely they are to spread it, for repeated exposure leads to an increased belief of this information~\citep{guess2020fake}. Ultimately, healthy democracies depend on a well-informed public, the very possibility of which is undermined by this new deluge of digital misinformation~\citep{lewandowsky2012misinformation}.

In response to these challenges, we present the results of a small pilot study of a media literacy campaign in Indonesia. This pilot study is the first step in a long-running program to understand the effects of digital misinformation on individuals in developing digital economies -- those who may not be sufficiently equipped to understand and navigate perils that proliferate on social media platforms.

There are many active projects seeking effective ways to stop this spread of misinformation, from new AI-based misinformation detection warning systems~\citep{yankoski2020ai}, to ``inoculation'' style misinformation games such as Harmony Square~\citep{roozenbeek2020breaking}, to fact checking organizations that meticulously comb through available evidence to affirm or debunk claims, such as Snopes.com. But there is no silver bullet for this accelerating problem: the adversarial nature of misinformation content creation processes means that AI-based misinformation detection systems will likely encourage the development of more sophisticated misinformation campaigns designed to elude the latest generation of AI-based detection systems\citep{anderson2017future}. Furthermore, fact-checking efforts may backfire by actually increasing the visibility of the very falsehoods they are seeking to debunk. For even if exposure to misinformation comes by seeing a piece of misinformation proven false, once a person has been exposed it is very difficult to turn them back to a state where they have never heard of it~\citep{lewandowsky2012misinformation}.

Many view media literacy and fact checking as the obvious solution(s) to misinformation and fake news~\citep{kim2020leveraging}. For example, watching others be corrected on social media can reduce beliefs in misperceptions~\citep{bode2020right}, and techniques such as gamification~\cite{chang2020news}, inoculation theory, and prebunking can reduce a user’s susceptibility to misinformation~\cite{roozenbeek2020breaking,lee2018fake}. Indeed, social media platforms like Facebook and Twitter are rolling out fact checking systems~\citep{fowler2020twitter} and Google is promoting its Be Internet Awesome media literacy program~\citep{seale2018internet}. However, others view these efforts with warranted suspicion. danah boyd~\citeyearpar{boyd2017did} argues that ``thorny problems of fake news and the spread of conspiracy theories have, in part, origins in efforts to educate people against misinformation'' because media literacy campaigns naturally ask the savvy information consumer to question the narratives that are presented to them. A study by \cite{craft2017news} showed that higher levels of news literacy predicted a lower endorsement of conspiracy theories. In contrast,~\cite{jones2021does} found that only information literacy, which was measured by a skills test, predicted a user’s ability to recognize fake news stories; whereas a user’s media, news, and digital literacy, which are self-reported metrics, were not predictive of a user’s ability to recognize fake news stories. 

In this article we offer preliminary results from our pilot study in combating the growing phenomenon of online misinformation: a social media literacy education campaign strategy that seeks to empower new digital arrivals in the Republic of Indonesia with increased ability to identify misinformation. After an assessment of the media landscape in Indonesia, we created 6 animated videos and 2 live-action videos demonstrating various lessons in online social media literacy. These lessons were presented as advertisements on YouTube, Google, Facebook, and Twitter and backed by \url{http://literasimediasosial.id}. Building off of other work such as Harmony Square~\citep{roozenbeek2020breaking} and Bad News~\citep{basol2020good}, which are each games in which players learn how misinformation spreads online, we ask the following research question: Does online media literacy content lead to an increased ability to identify misinformation? In other words, can we teach people to do their own fact checking, which then would result in a potentially scalable solution to solve the global problem of misinformation? We measure the effectiveness of our campaign by asking visitors to our website via a phone survey to rate the accuracy of true, misleading, and false headlines. The results from this pilot study suggest that there is a modest increase in the ability to determine whether a story is real or not after engaging our media literacy lessons. 

\subsection*{Why Indonesia?}
Indonesia is a large democracy with a rapidly growing Internet user base and social media penetration. As of 2019, approximately 68\% of Indonesia’s 270 million citizens were online, a figure that represents a dramatic increase from the approximately 43\% of the population just four years prior in 2015. By 2025 an estimated 89\% of Indonesians will be online. As of 2019, 88\%, 84\%, 82\%, and 79\% of Indonesian internet users self-reported use of YouTube, WhatsApp, Facebook, and Instagram, respectively~\citep{statista2020}. Because of this, many Indonesian citizens are new to the Internet. We position our work in this country to measure the effectiveness of this media literacy approach to a population which has had less exposure to the Internet, as compared to a Western audience. Additionally, this work provides important insights into the problems of online misinformation and propaganda, specifically within Southeast Asia, which remains an understudied region. Specifically, this context allows us to perform our research in a geographic area which is both in the Muslim world and in a country that is subjected to the Chinese sphere of influence. While similar previous work, such as Harmony Square and Bad News Game, was done targeting a Western audience, this study specifically targets another cultural context.

Many political events are frequently met with substantial coordinated misinformation campaigns (known as ``hoaxes'' in the Indonesian context). For example, in the 2018 Jakarta mayoral election, recent reports have indicated that the election campaigns paid as much as 280 USD per month to individuals who would promote messages from a particular candidate on social media~\cite{lamb2018felt}. Additionally, digital misinformation played a significant and highly divisive role in the national election in April of 2019~\citep{theisen2021automatic}. Protests against the results of the election resulted in six deaths and the temporary suspension of access to social media platforms by the Indonesian government~\citep{bbc2021}.

As more Indonesians gain reliable access to the Internet and begin to use online social media platforms, it is possible that those who are new to the Internet have not yet fully developed the media literacy skills needed to distinguish between trustworthy and false news sources, and may therefore be vulnerable to manipulation through misinformation. Because Indonesia is a relatively young democracy, traditional democratic institutions like the press may not be as robust as those institutions in more established democracies~\citep{bennett2018disinformation}.  Furthermore, the susceptibility of the voting population to misinformation campaigns may also pose a threat to the stability of the democratic institutions of the nation itself.

\subsection*{Assessing the Online and Social Media Landscape in Indonesia}

Our initial work focused on identifying legitimate news stories, propaganda, and disinformation, and the popular narratives and hashtags that were used in these stories and across these domains. For this we used a methodology developed by Moonshot, a company which uses social media monitoring, digital campaigns, and interventions to counter online harms.

This methodology uses a combination of desk research, consultation and workshops with subject-matter experts, open-source intelligence methods, and computer systems to identify words, phrases, tropes, slogans, memes, slang, and other indicators of engagement, in multiple languages and across search, social media, image boards, forums, apps, and other discursive online spaces as necessary. Exact sources vary depending on the subject matter and for this deployment in February 2019 we focused on words and phrases indicative of intent to engage with disinformation on Google Search, YouTube, Facebook, and Twitter.

The fundamental output of this methodology is a clear understanding of how interest in a given online harm manifests online. Depending on the exact methods of collection, this can be broken down by time, platform, aggregate user location, age and gender, subcategories of harm and different levels of risk. This understanding is then used to inform the creation of campaigns designed to reduce the impact of that harm, as was the case for this project. 

Analysis of the resulting dataset revealed four common themes of disinformation specific to the Indonesian context: (1) anti-Chinese, (2) anti-Communist, (3) Islamic chauvinism, and (4) political smears. This assessment of the social media and search engine landscape in Indonesia was used to inform the specific content used in the social media literacy campaign, based upon the common themes of disinformation that were identified.

\section*{Methods}
\subsection*{Social Media Literacy Intervention}

We developed a preliminary social media literacy education website (\url{http://literasimediasosial.id}) consisting of six informative lessons with several short educational videos and context-specific slogans that encourage social media behavior characterized by an increased ability to identify misinformation for widespread online delivery. This Learn To Discern (L2D)~\citep{murrock2018winning} approach, developed by IREX, builds communities' resilience to state-sponsored misinformation, inoculates communities against public health misinformation, promotes inclusive communities by empowering its members to recognize and reject divisive narratives and hate speech, improves young people’s ability to navigate increasingly polluted online spaces, and enables leaders to shape decisions based on facts and quality information~\citep{vogt2021learn}. L2D is a purely demand-driven approach to media literacy, encouraging participants to increase self-awareness of their own media environments and the forces or factors that affect the news and information that they consume. L2D is traditionally presented as a training program and it typically includes a package of in-person activities, online games, distance-learning courses, public service announcements, and other methods that are tailored to the needs of social media users. Examples of L2D lessons are shown in Fig.~\ref{fig:ad_impression}. For this campaign several short explainer videos were created. The full-length videos are hosted on YouTube (\url{https://www.youtube.com/channel/UCU7jNlA-4gH3cxsFV6_xkjw}) and direct users to \url{http://literasimediasosial.id}. Shortened versions of these animated and live action videos were also created to capture various aspects of the media literacy curriculum in short commercial-length snippets.

\subsection*{Delivering Media Literacy Content to Social Media Users}

Whereas previous media literacy efforts have been shown to be effective when conducted via in-person classroom settings~\citep{hobbs2003measuring}, our goal was to use the digitized lessons and explainer videos to directly reach users – not just students – within the social media platform itself. The Redirect Method~\citep{helmus2018assessing} is a methodology created by Moonshot and Jigsaw in 2016, originally to fight against violent extremism. It is now used by Moonshot to counter a range of online harms by reaching individuals who perform searches which indicate intent to engage in, or that they are affected by, harmful content on social media or on search engines, with positive, alternative content.

Just as other companies use advertisements on social media and search engines to sell material products to an audience defined by, in whole or in part, the keywords they are searching for (for instance, a vacuum cleaner company might bid for advertising space from users who are searching for cleaning products), the Redirect Method places ads in the search results and social media feeds of users who are searching for pre-identified terms that we have associated with disinformation.

The Redirect Method can be extensively tailored to platform requirements and campaign goals, but at its core are three fundamental components: the indicators of risk (e.g. keywords); the  9 advertisements triggered by the indicators; and the content to which users are redirected by the advertisements.

\subsection*{The Indicators of Risk}

We curated an extensive gazette of keywords and phrases indicative of a desire to engage with harmful content -- in this case, disinformation. The keywords for this campaign were created through a combination of desk research, consultation, and workshops with independent and local subject-matter experts and fact-checking organizations (some of whom shared their own keywords), and continuous mining for new keywords throughout the course of the project. The database currently contains more than 1,000 individual keyword phrases in Bahasa Indonesia.

Of the keywords identified for the campaign as indicative of intent to engage with disinformation, the three most-searched for were as follows:

\begin{enumerate}
    \item ``jokowi pki'', which translates to, “Jokowi [President Joko Widodo] is a member of PKI [the communist party of Indonesia]”
    \item ``PDIP adalah topeng pki'', which translates to, ``PDIP [the Indonesian Democratic Party of Struggle and the party to which the President belongs] is a mask of the PKI''
    \item ``9 naga pendukung jokowi'', which translates to ``the nine dragons support Jokowi''. The nine dragons refers to an age-old conspiracy that the capital city of Indonesia and its politicians are controlled by an underworld of nine Chinese mafia bosses. It is a myth based on the racist trope of ``ethnic-Chinese control'', used to instill fear and distrust.
\end{enumerate}

\subsection*{The Advertisements}

\begin{figure}
    \centering
    \includegraphics[width=.5\textwidth]{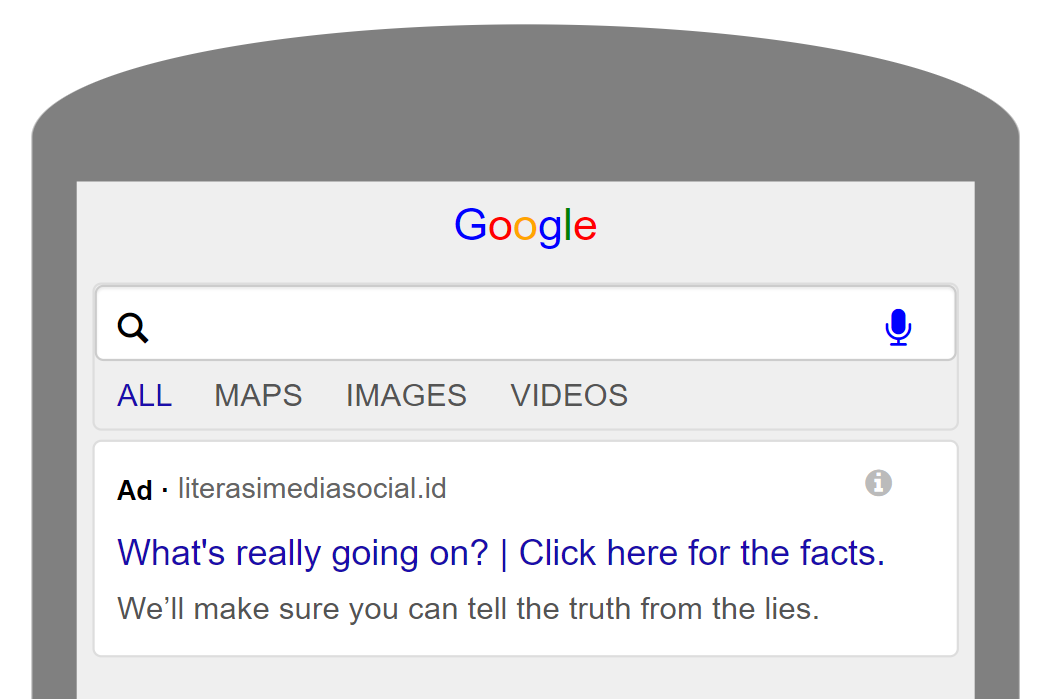}
    \caption{An English-language version of one of the advertisements used in this deployment of
the Redirect Method, visible only to users who were otherwise searching for disinformation
(actual ad text was in Bahasa Indonesia).}
    \label{fig:advert}
\end{figure}

The fundamental purpose of these advertisements is the same as those used in the commercial
sector, \ie, to entice people to click on them. The difference in the case of this deployment of
the Redirect Method is that social gain takes the place of commercial gain, with users who were
initially seeking disinformation being taken instead to content designed to improve their media
literacy

\begin{figure}
    \centering
    \includegraphics[width=.5\textwidth]{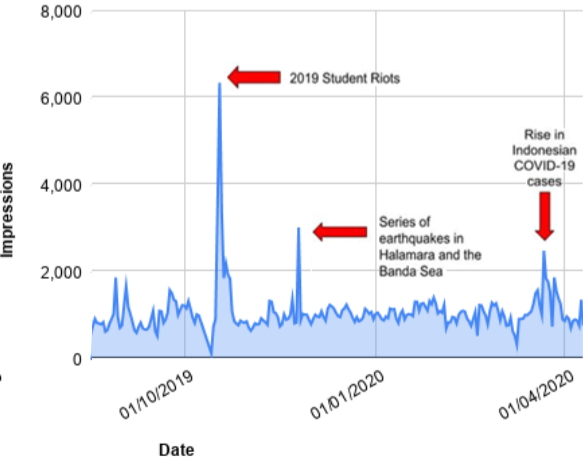}
    \caption{Ad impressions on Google Search for disinformation content. Disinformation-related
searches in Indonesia peaked during significant political and social events, such as the 2019
student riots, and the beginning of the COVID-19 pandemic.}
    \label{fig:ad_impression}
\end{figure}

Using Google Ads to display advertisements, \eg, Fig.~\ref{fig:advert}, in response to disinformation-related keyword searches means we can gather data on the number of times our advertisements were triggered (`impressions'), thereby gaining an insight into the volume of searches for disinformation. When plotted by time, as in Fig 2., we can also check for correlations in the search data with relevant and notable offline events.

\subsection*{The Content}

\begin{figure}
    \centering
    \includegraphics[width=.6\textwidth]{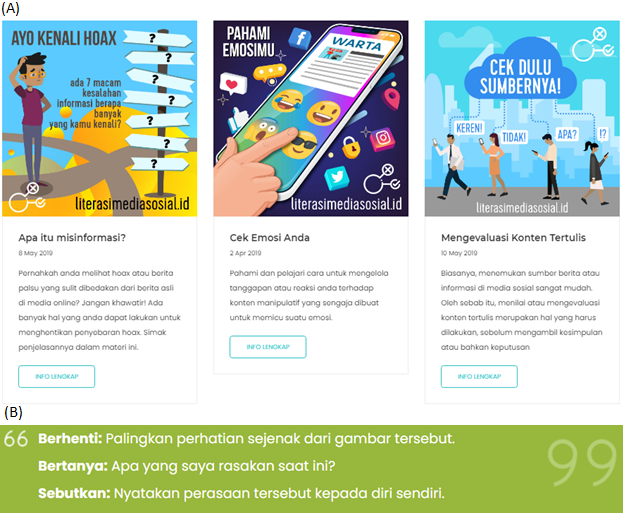}
    \caption{(A) Sample of media literacy lessons. These lessons translate as: What is
misinformation?, Check Your Emotions, and Evaluating Written Content. Each lesson contained
a short video and simple phrases which promote responsible social media behavior. (B) Snippet
from the lesson on ``Checking your Emotions.'' The translation is: ``Stop: Take a moment to pay
attention to the image; Ask: How am I feeling right now?; State it: Express these feelings to
yourself.'' We found that many of the visitors were spending less than a minute on our website,
so it was necessary to condense the lessons into short, concise bullet points and present them
like an advertisement.}
    \label{fig:lesson}
\end{figure}

Our social media literacy and search engine campaign was launched in August 2019 on Facebook, Twitter, Instagram, Google, and YouTube and concluded in April 2020. In total, the campaign generated 3.4 million impressions approximately equally distributed over 7 different video lessons. Those impressions resulted in 72,976 unique page views of the campaign website. Because of the wide reach of the advertisement campaign, we cannot be certain that individuals in our control group did not see our content. However, because 72,976 unique user sessions page views represents approximately 0.03\% of Indonesia’s estimated 183 million online users, we feel fairly confident that most of the control group users did not see our media literacy content, though we do not rule out this possibility. A sample of the media literacy lessons is illustrated in Fig.~\ref{fig:lesson}. 

Overall, we received 3,444,398.2 impressions (as calculated by each platform). Of those, 72,976 users clicked through to the website. In Table 1, we show how long each user spent on our site based upon the platform that the user was coming from. Additional campaign statistics are described in the Table~\ref{tab:reach} in the Appendix.

\subsection*{Phone Surveys}
After the conclusion of our social media literacy campaign, a team of interviewers based in Jakarta, Indonesia, conducted a computer-assisted telephone interview (CATI) for 1,000 successful interviews. The surveys were nationally representative proportionate to 2019 census data estimates at the province-level of location, age group, and gender. The main interview language of the survey was Bahasa Indonesia, but the interviewer team was able to switch to other languages such as Balinese and Javanese if requested by the interviewee.

The research protocol and questionnaire were approved by the internal ethics committee at the University of Notre Dame (\#18-11-5009). Data is anonymized, but demographic information is included. Data and full cross tabulations can be found at \url{https://www.geopoll.com/misinformation-indonesia/}

Visitors to the media literacy education website were asked to be included in a phone survey in exchange for approximately \$.50 USD equivalent in local currency in phone credit. 331 individuals agreed and provided their phone number, and of those, 94 completed the CATI survey. This constitutes the treatment group. The control group was comprised of 996 respondents to verified random digit dialing phone surveys. 

The treatment group was both older and skewed male as compared to the control, as shown in Tables A2 and A3 in the Appendix. We asked participants about general demographic information, inquired about their media consumption habits, and also queried their ability to correctly judge the veracity of news headlines that we presented to them. We acknowledge that the treatment group is likely to have self-selection bias. In addition, users who were redirected to the website were targeted because of their propensity to search for misinformation subjects, which presents another avenue for data bias. However, we believe that as a pilot study, this data provides important information about the efficacy of media literacy campaigns which invites further study. During data collection the survey team made approximately 4500 phone calls that resulted in a 21\% response rate. The margin of error for the survey at the 95\% confidence level is +/- 3.10\%.

\section*{Findings}

To judge the effectiveness of the social media literacy campaigns, each survey participant was asked if they recognized two true, one misleading, and one false headline. If so, we asked them to rate the accuracy of that headline on a Likert scale from very accurate to very inaccurate. Our goal was to examine the differences between how the control and treatment groups rated the accuracy of headlines. An improvement in the treatment group’s ability to correctly identify false, misleading, and real stories when compared to the control group would demonstrate that our campaign had an effect on a user’s ability to analyze the news headlines they encountered. 

\vspace{.5cm}
\noindent\textbf{Finding 1}: Users who encountered the media literacy content were more likely to identify false news than those that did not.
\vspace{.2cm}

\begin{figure}
    \centering
    \includegraphics[width=.7\textwidth]{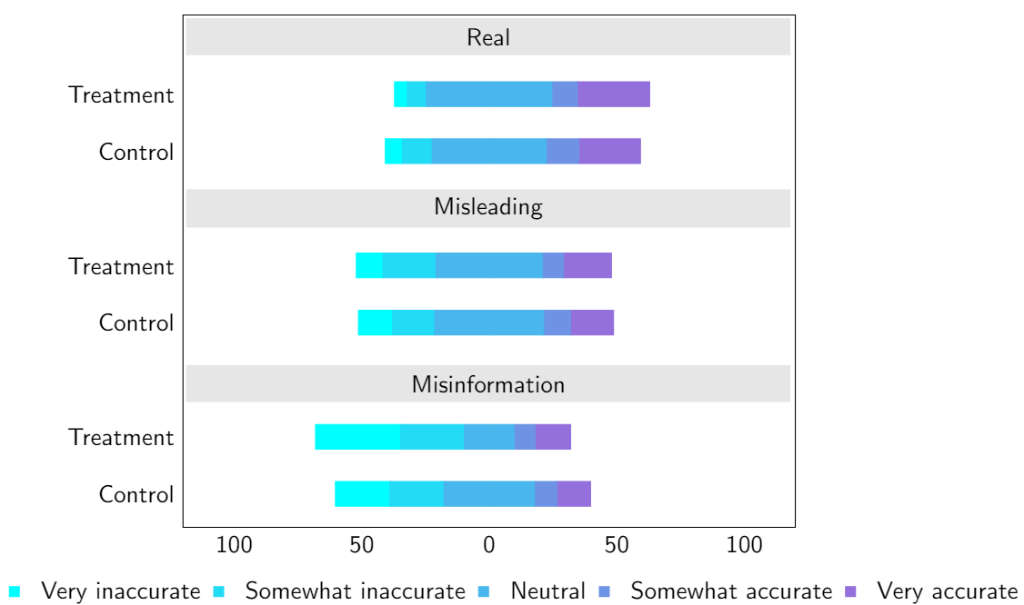}
    \caption{58.3 percent of our treatment group was able to identify the misinformation as either very or somewhat inaccurate, compared with 42.4 percent of the control group. We supply the number of respondents that gave an accuracy rating for each headline. For this plot, we averaged together the number of individuals who replied to our real headlines, since each person was given two real headlines to rate.}
    \label{fig:result1}
\end{figure}

\begin{table}[t]
    \centering
    \small{
    \begin{tabular}{@{}lrrrrrr@{}}
        \toprule
        \textbf{Headline} & \textbf{Group} & $N$ & \textbf{Mean Rank} & $\rho$ & \textbf{Significance} & \textbf{Remarks}  \\
         \midrule
        \multirow{2}{*}{\textbf{Real}} & Control   & 1113 & 3.36 & \multirow{2}{*}{0.470} & \multirow{2}{*}{0.871} & \multirow{2}{*}{Not Sig.} \\
                              & Treatment & 120  & 3.49 & & \\
        \multirow{2}{*}{\textbf{Misleading}} & Control   & 321 & 3.00 & \multirow{2}{*}{0.497} & \multirow{2}{*}{0.520} & \multirow{2}{*}{Not Sig.} \\
                              & Treatment & 48  & 3.04 & & \\
        \multirow{2}{*}{\textbf{False}} & Control   & 273 & 2.71 & \multirow{2}{*}{0.569} & \multirow{2}{*}{0.081} & \multirow{2}{*}{Sig. at 90\% confidence} \\
                              & Treatment & 36  & 2.44 & & \\                          
        \bottomrule
    \end{tabular}
    }
    \caption{Treatment users were statistically better at identifying misinformation headlines
compared to the control group at 90\% confidence.}
    \label{tab:headlineresults}
\end{table}

Our preliminary findings from this pilot study suggest that our media literacy intervention was positively correlated to a respondent’s ability to identify misinformation. Treatment group users in our pilot study were 15.9\% more likely to identify a misinformation headline as either very or somewhat inaccurate. These results are significant at 90 percent confidence (Mann–Whitney $\rho$ = 0.569, $N$ = 309, $p$ = 0.081 one-tailed). Figure~\ref{fig:result1} shows the results of the Likert-scale responses and Table 2 shows the statistical breakdown of the responses. Because the treatment group consisted of individuals who were actively searching for misinformation, these results are even more encouraging because they suggest that our campaign had a measurable effect even on those who are familiar with and are actively searching for misinformation content. We plan to expand upon this pilot study’s encouraging results in our future research, as we acknowledge the limitations of such a small sample size.

Across both the control and treatment groups, accuracy ratings were very similar for the different headline types. Interestingly, as shown in Fig.~\ref{fig:result1}, it appears that relatively few individuals in the treatment group had a neutral opinion of the misinformation story, as compared to the other groups and headline types. This suggests that our media literacy campaign might have persuaded those who were not sure about the validity of the story that the narrative was false, via the new skills they acquired from the website. This is supported by the observation that the truthfulness rating of the misinformation headlines was similar between the treatment and control groups. More research is needed to understand the effect of media literacy campaigns on these fence-sitters -– those who have neutral feelings about the headlines that they see, and are unable to tell if these stories are true or false. 

In summary, we found that our social media literacy intervention is associated with an increase in the ability of users to correctly identify misinformation. However, despite these findings, participants in our social media literacy campaign did not show an increased ability to accurately identify misleading or real news stories. This phenomenon presents an intriguing area for future research. 

\vspace{.5cm}
\noindent\textbf{Finding 2}: False news did not spread as broadly in Indonesia as real news.
\vspace{.2cm}

\begin{table}[t]
    \centering
    \small{
    \begin{tabular}{@{}lrrrr@{}}
        \toprule
        \textbf{Heard of Story?} & \textbf{Real ($N$ = 1865)} & \textbf{Misleading ($N$ = 875)} & \textbf{False ($N$ = 870)} & \textbf{Total}  \\
         \midrule
        \textbf{Yes} & 1332 (71.4\%) & 401 (45.8\%) &  333 (38.2\%) &  $N$ = 2066 \\
        \textbf{No} & 533 (28.5\%) &   474 (54.1\%) &  537 (61.7\%) &  $N$ = 1544 \\
        \bottomrule
    \end{tabular}
    }
    \caption{Awareness of news stories is positively correlated with its veracity $\chi^2$(2, $N$ = 3,610) = 327.60 $p<$ 0.001}
    \label{tab:heard}
\end{table}

Interestingly and rather unexpectedly, we found that the headline accuracy was positively correlated with its reach. These results, shown in Table~\ref{tab:heard}, suggest that misinformation does not spread quite as easily as real news stories in Indonesia – however, further research is needed to understand this phenomenon. In future work, we plan to expand the number of headlines that we supply to our survey respondents.

Our results suggest that our media literacy campaign had an effect on the ability of individuals to accurately label hoaxes as misinformation. However, further research is needed to answer some of the questions that our findings raise. For instance, we see that large numbers of respondents have neutral feelings about news stories. While our campaign focused on identifying misinformation, perhaps future work can promote trust in legitimate news sources, which would create statistically significant differences in the abilities of media literacy campaign participants to identify real headlines as such, not simply to identify misinformation.

\section*{Conclusions}
In this work, we present findings from our early pilot study on the efficacy of a media literacy campaign in Indonesia in helping users identify misinformation headlines. These results indicate that 58.3 percent of users who visited our media literacy website and engaged with the education content were able to correctly identify misinformation headlines as inaccurate, compared to our 42.4 percent of the control group. These results suggest that it is possible to use a media literacy approach to help citizens spot and identify misinformation.

This work presented a small, limited study to understand the misinformation landscape in Indonesia, and demonstrates the feasibility of conducting a larger research project of this nature. Since our results were positive, we plan on expanding this study in several ways. First, we will introduce a gamification approach - analogous to Harmony Square and Bad News game -- into our media literacy education materials. Second, we will enlarge the sample size and test the significance of the results on larger groups. Third, we will track information such as how long a user stayed on the site and engaged with the material. Fourth, we will ask more questions about Internet usage, such as how long a user has been using the Internet. Fifth, we will ask users about more disinformation headlines, since currently we only ask them about the veracity of one headline. With these additional considerations, we will create and deploy a broader study which will give us even more insights into the efficacy of media literacy campaigns.

We present these preliminary results to others that might be considering larger, more comprehensive studies in this area. We hope that our work will help those who are designing such studies, and implement the lessons learned from our study to further study media literacy, especially in developing countries or for new digital arrivals.

\section*{Acknowledgements} We would like to acknowledge and thank Carolina Rocha da Silva, Rachel Fielden, Joel Turner, Tavian MacKinnon, and Walter Scheirer, Joshua Macleder, and Anders Mantius for their
assistance on this project. The funding for this research study was supported by USAID Cooperative Agreement \#7200AA18CA00059


\newpage

\appendix

\setcounter{table}{0}
\renewcommand{\thetable}{A\arabic{table}}

\begin{table}[t]
    \centering
    \small{
    \begin{tabular}{@{}lrrrr@{}}
        \toprule
        \textbf{Platform} & \textbf{Impressions} & \textbf{Clicks} & \textbf{Avg. Duration (sec)} & \textbf{Click Through Rate}  \\
         \midrule
        \textbf{Google} & 252,511.0    & 6,877  & --    & 4.0\% \\
        \textbf{Facebook} & 568,452.3  & 22,747 & 14.17 & 4.0\% \\
        \textbf{Instagram} & 525,292.8 & 2,315  & 27.54 & 0.4\% \\
        \textbf{Twitter} & 294,793.4   & 2,265  & 35.20 & 0.8\% \\
        \textbf{Youtube} & 1,803,348.8 & 38,772 & 0.62  & 2.2\% \\
        \bottomrule
    \end{tabular}
    }
    \caption{Reach and viewership statistics of media literacy campaign}
    \label{tab:reach}
\end{table}

\begin{table}[t]
    \centering
    \small{
    \begin{tabular}{@{}lrrr@{}}
        \toprule
        \textbf{Age} & \textbf{Treatment} ($N$ = 94) &  \textbf{Control} ($N$ = 906) & \textbf{Total} ($N$ = 1000)  \\
         \midrule
        15-24 & 13 (13.8\%) & 201 (22.1\%) & $N$ = 214 \\
        25-35 & 23 (24.4\%) & 305 (33.6\%) & $N$ = 328 \\
        36-45 & 18 (19.1\%) & 216 (23.8\%) & $N$ = 234 \\
        46-55 & 21 (22.3\%) & 123 (13.5\%) & $N$ = 144 \\
        56-65 & 15 (15.9\%) & 48 (5.2\%) & $N$ = 63 \\
        66+   & 4 (4.2\%)   & 13 (1.4\%) & $N$ = 17 \\
        \bottomrule
    \end{tabular}
    }
    \caption{Treatment group is older than our control group $\chi^2$(5, $N$ = 1,000) = 29.66 $p<$ 0.001)}
    \label{tab:older}
\end{table}

\begin{table}[t]
    \centering
    \small{
    \begin{tabular}{@{}lrrr@{}}
        \toprule
        \textbf{Gender} & \textbf{Treatment} ($N$ = 94) &  \textbf{Control} ($N$ = 906) & \textbf{Total} ($N$ = 1000)  \\
         \midrule
        MALE   & 72 (76.5\%) & 428 (47.2\%) &  $N$ = 500 \\
        FEMALE & 22 (23.4\%) & 478 (52.7\%) &  $N$ = 500 \\
        \bottomrule
    \end{tabular}
    }
    \caption{Treatment group is more male than our control group $\chi^2$(1, $N$ = 1,000) = 29.35 $p<$
0.001)}
    \label{tab:male}
\end{table}

\begin{table}[t]
    \centering
    \small{
    \begin{tabular}{@{}lrrr@{}}
        \toprule
        \textbf{Urban/Rural?} & \textbf{Treatment} ($N$ = 93) &  \textbf{Control} ($N$ = 901) & \textbf{Total} ($N$ = 994)  \\
         \midrule
        Urban & 46 (49.4\%) & 404 (44.8\%) & $N$ = 450\\
        Rural & 47 (50.5\%) & 497 (55.1\%) & $N$ = 544\\
        \bottomrule
    \end{tabular}
    }
    \caption{Treatment group is similar to our control group based upon urban/rural identity $\chi^2$(1, $N$ = 994) = 0.727 $p$ = 0.393)}
    \label{tab:urban}
\end{table}

\begin{table}[t]
    \centering
    \small{
    \begin{tabular}{@{}lrrr@{}}
        \toprule
        \textbf{Religion} & \textbf{Treatment} ($N$ = 91) &  \textbf{Control} ($N$ = 889) & \textbf{Total} ($N$ = 990)  \\
         \midrule
        Muslim    & 87 (95.6\%) & 827 (91.1\%) & $N$ = 914 \\
        Christian & 4 (4.3\%) & 58 (6.4\%) & $N$ = 62 \\
        Other     & 0 (0.0\%) & 13 (1.4\%) & $N$ = 13 \\
        None      & 0 (0.0\%) & 1 (0.1\%) & $N$ = 1 \\
        \bottomrule
    \end{tabular}
    }
    \caption{Treatment group is similar to our control group based upon religious identity $\chi^2$(3, $N$ = 990) = 2.09 $p$ = 0.553) }
    \label{tab:religion}
\end{table}

\end{document}